\documentclass[twocolumn,showpacs,preprintnumbers,amsmath,amssymb,nofootinbib]{revtex4-1}

\usepackage{graphicx}
\usepackage{dcolumn}
\usepackage{bm}
\usepackage{amsfonts}

\def\bequ{\begin{equation}}
\def\eequ{\end{equation}}
\def\be{\begin{equation}}
\def\ee{\end{equation}}

\begin{document}

\title{Dirac perturbations on Schwarzschild-Anti-de Sitter spacetimes: \\Generic boundary conditions and new quasinormal modes}

\author{Mengjie Wang$^1$}
\email{mjwang@hunnu.edu.cn}
\author{Carlos Herdeiro$^2$}
\email{herdeiro@ua.pt}
\author{Jiliang Jing$^1$}
\email{jljing@hunnu.edu.cn}
\affiliation{\vspace{2mm}
$^1$Department of Physics, Key Laboratory of Low Dimensional Quantum Structures and
Quantum Control of Ministry of Education, and Synergetic Innovation Center for Quantum Effects and Applications, Hunan Normal University,
Changsha, Hunan 410081, P.R. China
\vspace{1.3mm}
\\
$^2$Departamento de F\'\i sica da Universidade de Aveiro and CIDMA, Campus de Santiago, 3810-183 Aveiro, Portugal \vspace{1mm}}%

\date{\today}

\begin{abstract}
We study Dirac quasinormal modes of Schwarzschild-Anti-de Sitter (Schwarzschild-AdS) black holes, following the generic principle for allowed boundary conditions proposed in \cite{PhysRevD.92.124006}. After deriving the equations of motion for Dirac fields on the aforementioned background, we impose \textit{vanishing energy flux} boundary conditions to solve these equations. We find a set of \textit{two} Robin boundary conditions are allowed. These two boundary conditions are used to calculate Dirac normal modes on empty AdS and quasinormal modes on Schwarzschild-AdS black holes. In the former case, we recover the known normal modes of empty AdS; in the latter case, the two sets of Robin boundary conditions lead to two different branches of quasinormal modes. The impact on these modes of the black hole size, the angular momentum quantum number and the overtone number are discussed. Our results show that vanishing energy flux boundary conditions are a robust principle, applicable not only to bosonic fields but also to fermionic fields.
\end{abstract}

\maketitle

\section{Introduction}

Black holes (BHs) are often claimed to be the simplest macroscopic bodies in Nature. This view arises from the  theoretical paradigm that BHs can be uniquely characterized by their mass, spin and charge -- the \textit{no-hair conjecture}~\cite{Misner:1974qy}.  The first direct observations of gravitational waves, recently reported~\cite{Abbott:2016blz,Abbott:2016nmj,Abbott:2017vtc,Abbott:2017oio}, together with many other observations in the electromagnetic channel, are opening a new era in testing strong gravity~\cite{Berti:2015itd}, and will, in time, provide evidence for, or against, this paradigm.\footnote{Theoretically, there are many counter-examples to the no-hair conjecture, see $e.g.$ the reviews~\cite{Herdeiro:2015waa,Volkov:2016ehx}, including some hairy BHs continuously connected to the Kerr solution~\cite{Herdeiro:2014goa,Herdeiro:2016tmi} that may form dynamically~\cite{East:2017ovw,Herdeiro:2017phl}.}

The aforementioned conceptual simplicity contrasts with the technical complexity of BH physics. BHs are non-linear solutions of a highly non-linear theory, and studying their dynamical aspects is often a formidable challenge. In this respect, perturbative methods are an important complement to the large infratructure nonlinear numerics, and stand out as a useful tool in studying the interactions between BHs and fundamental test fields. Since the celebrated work by Teukolsky~\cite{Teukolsky:1973ha}, perturbation equations for different spin fields have been obtained, which provide the foundations to study various dynamical aspects, such as quasinormal modes and quasi-bound states. The former case is particularly interesting -- see, $e.g.$, the reviews~\cite{Kokkotas:1999bd,Berti:2009kk,Konoplya:2011qq} and references therein -- , since in asymptotically flat spacetimes it can be used to test strong gravity in the gravitational wave era, while in asymptotically AdS spacetimes it can be used to obtain the timescale for the approach to thermal equilibrium.

Quasinormal modes of different spin fields on Schwarzschild-AdS spacetimes have been studied extensively,  the scalar field being the most studied case~\cite{Chan:1996yk,Chan:1999sc,Horowitz:1999jd}. In the scalar case, the boundary condition taken requires the scalar field itself to vanish at the asymptotic boundary. This type of boundary condition was then generalized to study quasinormal modes for the Maxwell, gravitational and Dirac cases~\cite{Cardoso:2001bb,Cardoso:2003cj,Giammatteo:2004wp,Jing:2005ux}. As we have pointed out in~\cite{PhysRevD.92.124006}, however, by taking the Maxwell field as an example, this scalar-like boundary condition can not be applied to the Maxwell field, when using the Teukolsky formalism, which is central to separate perturbations on rotating BH backgrounds.\footnote{In the Regge-Wheeler formalism, the scalar-like boundary condition may miss one set of the modes.}

To overcome this issue, and gain a more transversal guiding principle, we have recently proposed a simple perspective on the boundary conditions for quasinormal modes in asymptotically AdS spacetimes. It follows the idea that the AdS boundary may be viewed as a perfect reflecting mirror in the sense that the energy flux vanishes at the asymptotic boundary. As we have shown explicitly by applying this principle to the Maxwell fields~\cite{PhysRevD.92.124006,Wang:2015fgp,Wang:2016dek,Wang:2016zci}, two families of boundary conditions are possible, yielding two branches of quasinormal modes. Moreover, the same two branches are obtained for both the Regge-Wheeler and the Teukolsky equations, showing the setup is consistent.

The physical principle we have proposed on boundary conditions originates from the asymptotic AdS structure, regardless of the spin of the perturbing field. As such, we shall initialize a systematic study on Dirac field perturbations in asymptotically AdS spacetimes, under vanishing energy flux boundary \mbox{conditions}. Dirac quasinormal modes on Schwarzschild-AdS BHs have been addressed in, $e.g.$,~\cite{Giammatteo:2004wp,Jing:2005ux}, wherein  scalar-like boundary condition has been imposed. Here, as the first paper of our study on Dirac fields with the new boundary conditions, we focus on a massless neutral Dirac field interacting with a Schwarzschild-AdS BH.

To setup our study, we first present the Dirac equations on Schwarzschild-AdS BHs, both by using the $\gamma$ matrices~\cite{Unruh:1973bda} method and by using the Teukolsky~\cite{Teukolsky:1973ha} approach. Requiring the energy flux to vanish at the asymptotic boundary, we then calculate the explicit boundary conditions associated with the Dirac equations in the $\gamma$ matrices formalism. Similarly to the Maxwell case~\cite{PhysRevD.92.124006}, we obtain two sets of boundary conditions. These boundary conditions are computed for both the $R_1$ and $R_2$ equations, where $R_1$ and $R_2$ are the radial variables describing the two degrees of freedom of the Dirac fields. The same quasinormal modes are obtained for both equations, by imposing the corresponding boundary conditions. Furthermore, we verify that the Dirac equations in the Teukolsky formalism are simply related with the counterpart equations in the $\gamma$ matrices formalism. Based on this observation, one may easily obtain the corresponding boundary conditions for Dirac fields in the Teukolsky formalism. As expected, the same quasinormal modes for Dirac fields may be obtained using the two different formalisms.

The structure of this paper is organized as follows. In Section~\ref{seceq} we introduce the Schwarzschild-AdS geometry and derive the corresponding Dirac equations in the $\gamma$ matrices formalism. In Section~\ref{secbc} we show how to obtain \textit{two} Robin boundary conditions for Dirac fields in the aforementioned background, satisfying the vanishing energy flux requirement, at the AdS boundary. In Section~\ref{secana} we solve Dirac equations \textit{analytically}, by applying the boundary conditions obtained in the previous section, yielding the Dirac normal modes on empty AdS and the quasinormal modes on small Schwarzschild-AdS BHs. These calculations show clearly how to employ \textit{two} boundary conditions to obtain \textit{two} different sets of quasinormal modes. Numerical methods and results are presented in Section~\ref{secnum}, to illustrate how the parameters $r_+$ (the BH size), $\ell$ (the angular momentum quantum number) and $N$ (the overtone number) affect the two sets of modes. Final remarks and conclusions are presented in the last section. Some technicalities on the Dirac equations in the Teukolsky formalism, as well as some considerations on the number current for Dirac fields are left to the Appendix.

\section{background geometry and field equations}
\label{seceq}
In this section, we briefly review basic properties of Schwarzschild-AdS BHs, and derive the equations of motion for test Dirac fields on this background geometry.

\subsection{Schwarzschild-AdS BHs}
The line element of a Schwarzschild-AdS BH can be written as (observe we shall use a $(+---)$ signature)
\begin{equation}
ds^2=\dfrac{\Delta_r}{r^2}dt^2-\dfrac{r^2}{\Delta_r}dr^2-r^2d\theta^2-r^2\sin^2\theta d\varphi^2 \;,\label{metric}
\end{equation}
with the metric function
\begin{eqnarray}
\Delta_r\equiv r^2\left(1+\frac{r^2}{L^2}\right)-2Mr\;,\label{metricfunc}
\end{eqnarray}
where $L$ is the AdS radius and $M$ is the mass parameter. The event horizon $r_+$ is determined as the largest root of $\Delta_r(r_+)=0$. For a given $r_+$ the mass parameter can be expressed as
\begin{equation}
M=\dfrac{r_+(L^2+r_+^2)}{2L^2}\;.\nonumber
\end{equation}
Then, the Hawking temperature can be written in terms of $r_+$
\begin{equation}
T_H=\dfrac{\kappa}{2\pi}=\dfrac{3r_+^2+L^2}{4\pi r_+L^2}\;.
\end{equation}

\subsection{Dirac equations in the $\gamma$ matrices formalism}
The equations of motion for a massless Dirac field on a Schwarzschild-AdS background can be obtained in various ways~\cite{Unruh:1973bda,Chandrasekhar:1976ap,Dolan:2015eua}. In this subsection, we derive the Dirac equations  in the $\gamma$ matrices formalism, by adapting Unruh's original work~\cite{Unruh:1973bda}, wherein the equations of motion for a massless Dirac field on a Kerr BH was derived.

A massless Dirac field obeys the equation
\begin{equation}
\gamma^\mu(\partial_\mu-\Gamma_\mu)\Psi=0\;,\label{DiraceqU1}
\end{equation}
where the $\gamma$ matrices are defined as
\begin{align}
&\gamma^t=\sqrt{\frac{r^2}{\Delta_r}}\gamma^0\;,\;\;\;\;\;\;\gamma^r=\sqrt{\frac{\Delta_r}{r^2}}\gamma^3\;,\nonumber\\
&\gamma^\theta=\frac{1}{r}\gamma^1\;,\;\;\;\;\;\;\;\;\;\;\;\;\gamma^\varphi=\frac{1}{r\sin\theta}\gamma^2\;,\label{gammams}
\end{align}
with the ordinary flat spacetime Dirac matrices $\gamma^i (i=0, 1, 2, 3)$ in the Bjorken-Drell representation~\cite{Bjorken:100769}. The spin connection is
\begin{equation}
\Gamma_\mu=-\frac{1}{8}(\gamma^a\gamma^b-\gamma^b\gamma^a)\Sigma_{ab\mu}\;,\label{spincon}
\end{equation}
with
\begin{equation}
\Sigma_{ab\mu}=e_a^\nu(\partial_\mu e_{b\nu}-\Gamma^\alpha_{\nu\mu}e_{b\alpha})\;.\nonumber
\end{equation}
Letting
\begin{equation}
\Psi=\left(
\begin{matrix}
\eta\\
\eta
\end{matrix}
\right)\;,\label{psidecom1}
\end{equation}
with the ansatz
\begin{equation}
\eta=\frac{e^{-i\omega t}e^{im\varphi}}{(\Delta_rr^2\sin^2\theta)^{1/4}}
\left(
\begin{matrix}
R_1(r)S_1(\theta)\\
R_2(r)S_2(\theta)
\end{matrix}
\right)\;,\label{psidecom2}
\end{equation}
then Eq.~~\eqref{DiraceqU1} becomes a set of coupled first order equations
\begin{align}
&\Delta_r^{1/2}\left(\dfrac{d}{dr}-\dfrac{i\omega r^2}{\Delta_r}\right)R_1(r)=kR_2(r)\;,\label{firstorderr1}\\
&\Delta_r^{1/2}\left(\dfrac{d}{dr}+\dfrac{i\omega r^2}{\Delta_r}\right)R_2(r)=kR_1(r)\;,\label{firstorderr2}\\
&\left(\dfrac{d}{d\theta}-\dfrac{m}{\sin\theta}\right)S_1(\theta)=kS_2(\theta)\;,\label{firstorderang1}\\
&\left(\dfrac{d}{d\theta}+\dfrac{m}{\sin\theta}\right)S_2(\theta)=-kS_1(\theta)\;.\label{firstorderang2}
\end{align}
Second order equations can be easily obtained, from Eqs.~\eqref{firstorderr1}-\eqref{firstorderang2}; the radial part is
\begin{align}
&\Delta_r^{1/2}\dfrac{d}{dr}\left(\Delta_r^{1/2}\dfrac{d R_{1}}{dr}\right)+H_1(r)R_{1}=0\;,\label{DiracU2radialR1}\\
&\Delta_r^{1/2}\dfrac{d}{dr}\left(\Delta_r^{1/2}\dfrac{d R_{2}}{dr}\right)+H_2(r)R_{2}=0\;,\label{DiracU2radialR2}
\end{align}
with $R_1\equiv R_1(r), R_2\equiv R_2(r)$, and
\begin{align}
&H_1(r)=\dfrac{K_r^2+\tfrac{i}{2}K_r\Delta_r^\prime}{\Delta_r}-2i\omega r-k^2\;,\nonumber\\
&H_2(r)=\dfrac{K_r^2-\tfrac{i}{2}K_r\Delta_r^\prime}{\Delta_r}+2i\omega r-k^2\;,\nonumber
\end{align}
where $K_r = \omega r^2$, and the angular part is
\begin{align}
&\dfrac{d^2S_1}{d\theta^2}+\left(-\dfrac{m^2}{\sin^2\theta}+m\dfrac{\cos\theta}{\sin^2\theta}+k^2\right)S_{1}=0\;,\label{DiracU2angularS1}\\
&\dfrac{d^2S_2}{d\theta^2}+\left(-\dfrac{m^2}{\sin^2\theta}-m\dfrac{\cos\theta}{\sin^2\theta}+k^2\right)S_{2}=0\;,\label{DiracU2angularS2}
\end{align}
with $S_1\equiv S_1(\theta), S_2\equiv S_2(\theta)$. The solutions for these angular equations are spin-weighted spherical harmonics, and the corresponding eigenvalue is $k^2=(\ell+\frac{1}{2})^2$~\cite{Dolan:2009kj}.

The radial part of the second order differential equations, from Eq.~\eqref{DiracU2radialR1}-Eq.\eqref{DiracU2radialR2}, are what we are going to study in the remaining sections.

\section{boundary conditions}
\label{secbc}
To solve the differential equations ~\eqref{DiracU2radialR1}--\eqref{DiracU2radialR2}, one has to impose physically relevant boundary conditions. At the horizon, one imposes purely ingoing boundary conditions. At the asymptotic boundary, a scalar-like boundary condition is typically imposed~\cite{Giammatteo:2004wp,Jing:2005ux}. Here, however, we are going to study quasinormal modes for Dirac fields in Schwarzschild-AdS BHs by imposing vanishing energy flux boundary conditions, as proposed in~\cite{PhysRevD.92.124006,Wang:2015fgp,Wang:2016dek,Wang:2016zci}. This requirement follows the spirit that the AdS boundary may be regarded as a perfectly reflecting mirror in the sense that no flux can cross it.

Comparing with the scalar-like boundary condition, vanishing energy flux boundary conditions are applicable to
\begin{itemize}
\item[$\bullet$] the Dirac equations both in the $\gamma$ matrices formalism and in the Teukolsky formalism;
\item[$\bullet$] both the $R_1$ equation and the $R_2$ equation;
\end{itemize}
and one may obtain
\begin{itemize}
\item[$\bullet$] two different sets of explicit boundary conditions;
\item[$\bullet$] in particular, normal modes on empty AdS.
\end{itemize}

We start from the energy-momentum tensor for Dirac fields, which is defined as
\begin{equation}
T_{\mu \nu}=\dfrac{i}{8\pi}\bar{\Psi}\left[\gamma_\mu(\partial_\nu-\Gamma_\nu)+\gamma_\nu(\partial_\mu-\Gamma_\mu)\right]\Psi+c.c. \;,\label{EMTensorDirac}
\end{equation}
where $\bar{\Psi}\equiv \Psi^\dag\gamma^0$, and $c.c.$ stands for complex conjugate of the preceding terms. Note that $\gamma_\mu=g_{\mu\nu}\gamma^\nu$, where $\gamma^\nu$ is given in Eq.~\eqref{gammams}, the spin connection $\Gamma_\mu$ is given in Eq.~\eqref{spincon}, and $\Psi^\dag$ is the hermitian conjugate of $\Psi$.

To impose the boundary conditions we shall require, we take the definition of the energy flux through a 2-sphere at radial coordinate $r$:
\begin{equation}
\mathcal{F}|_r=\int_{S^2} \sin\theta d\theta d\varphi\; r^2 T^r_{\;\;t}\; ; \label{flux1}
\end{equation}
thus, we have to calculate $T^r_{\;\;t}$ firstly, which is given by
\begin{equation}
T^r_{\;\;t}=T^r_{\;\;t,\;\uppercase\expandafter{\romannumeral1}}+T^r_{\;\;t,\;\uppercase\expandafter{\romannumeral2}}\;,\nonumber\\
\end{equation}
with
\begin{equation}
T^r_{\;\;t,\;\uppercase\expandafter{\romannumeral1}}=\dfrac{\omega+\omega^{\ast}}{2\pi r^2\sin\theta}\left(|R_1|^2|S_1|^2-|R_2|^2|S_2|^2\right)\;,
\end{equation}
where $\omega^{\ast}$ is the complex conjugate of $\omega$, and $T^r_{\;\;t,\;\uppercase\expandafter{\romannumeral2}}$  vanishes after integrating over the sphere. Then, the energy flux becomes
\begin{equation}
\mathcal{F}|_r\propto\left(|R_1|^2-|R_2|^2\right)\;,\label{flux2}
\end{equation}
up to a factor independent of the radial coordinate, and where the angular functions $S_{\pm}(\theta)$ are normalized
\begin{equation}
\int_0^\pi d\theta\; |S_{1,2}(\theta)|^2=1\;.\nonumber
\end{equation}
\\
To obtain the asymptotic boundary condition for $R_{1}$, we make the asymptotic expansion from Eq.~\eqref{DiracU2radialR1}, and get
\begin{equation}
R_1 \sim \alpha_1+\beta_1\dfrac{L}{r}+\mathcal{O}\left(\dfrac{L^2}{r^2}\right)\;,\label{asysolR1}
\end{equation}
where $\alpha_1$ and $\beta_1$ are two integration constants.

With the relation between $R_1$ and $R_2$ in Eq.~\eqref{firstorderr1}, and making use of expansion for $R_1$ in Eq.~\eqref{asysolR1}, at infinity the energy flux in Eq.~\eqref{flux2} becomes
\begin{eqnarray}
\mathcal{F}|_{r,\infty}\propto k^2|\alpha_1|^2-|i\omega L\alpha_1+\beta_1|^2\;.\label{fluxinf}
\end{eqnarray}
Now we are able to impose \textit{energy flux vanishing boundary conditions}, i.e. $\mathcal{F}|_{r,\infty}=0$, which implies
\begin{equation}
k^2|\alpha_1|^2-|i\omega L\alpha_1+\beta_1|^2=0\;.
\end{equation}
It is easy to solve this quadratic equation and obtain the two solutions\footnote{Note that the relative phase between two moduli has been fixed by calculating normal modes. That is, for empty AdS normal modes are only allowed for this particular choice of phase.}
\begin{align}
\dfrac{\alpha_1}{\beta_1}=\dfrac{-i}{\ell+\frac{1}{2}-\omega L}\;,\label{bc1}\\
\dfrac{\alpha_1}{\beta_1}=\dfrac{i}{\ell+\frac{1}{2}+\omega L}\;,\label{bc2}
\end{align}
which tell us that the physical requirement of vanishing energy flux generates two sets of boundary conditions. This means that, in Schwarzschild-AdS BHs, there are two branches of quasinormal modes for Dirac fields, following the same logic as for the Maxwell case~\cite{PhysRevD.92.124006,Wang:2015fgp,Wang:2016dek,Wang:2016zci}.

We can also follow the same procedures to calculate the boundary conditions for $R_2$. As before, we first expand $R_2$ from Eq.~\eqref{DiracU2radialR2}, and get
\begin{equation}
R_{2} \sim \;\alpha_2+\beta_2\dfrac{L}{r}+\mathcal{O}\left(\dfrac{L^2}{r^2}\right)\;,\label{asysolR2}
\end{equation}
where $\alpha_2$ and $\beta_2$ are two integration constants. Then making use of the relation in Eq.~\eqref{firstorderr2}, Eq.~\eqref{flux2} gives the conditions
\begin{equation}
\dfrac{\alpha_2}{\beta_2}=\dfrac{i}{\ell+\frac{1}{2}-\omega L}\;,\;\;\;\dfrac{\alpha_2}{\beta_2}=\dfrac{-i}{\ell+\frac{1}{2}+\omega L}\;.\label{bcanother}
\end{equation}
Comparing boundary conditions for $R_1$, Eqs.~\eqref{bc1}--\eqref{bc2}, and for $R_2$, Eq.~\eqref{bcanother}, we notice that there is only a sign difference. As we have checked, solving the radial equation~\eqref{DiracU2radialR1} with the corresponding boundary conditions~\eqref{bc1},~\eqref{bc2} and the radial equation~\eqref{DiracU2radialR2} with the corresponding boundary conditions~\eqref{bcanother}, the same quasinormal frequencies are obtained. This implies that $R_1$ and $R_2$ encode the same information. Therefore, for concreteness, and without loss of generality, in the following we focus on the $R_1$ equation and the corresponding boundary conditions.

\section{Analytics}
\label{secana}
In this section, we solve Dirac equations \textit{analytically}, both to calculate normal modes on empty AdS and to calculate quasinormal modes on small Schwarzschild-AdS BHs. These calculations are performed in order to show how to employ the boundary conditions, Eqs.~\eqref{bc1}--\eqref{bc2}, to get two sets of modes.

\subsection{Dirac normal modes on empty AdS}

In an empty AdS spacetime (no BH), the radial Dirac equation~\eqref{DiracU2radialR1} keeps the same form, but with
\begin{equation}
\Delta_r = r^2 \left(1+\dfrac{r^2}{L^2}\right)\;.\label{AdSDelta}
\end{equation}
Then, the general solution of Eq.~\eqref{DiracU2radialR1},  with~\eqref{AdSDelta} is
\begin{eqnarray}
R_1&&=r^{\ell+\frac{1}{2}}(r-iL)^{\frac{\omega L}{2}}(r+iL)^{-\ell-\frac{1}{2}-\frac{\omega L}{2}}\Big[(-1)^{2\ell+1}2^{-2\ell-1}\nonumber\\&&\left(1+\dfrac{iL}{r}\right)^{2\ell+1}C_1F\Big(-\ell-\frac{1}{2},-\ell+\omega L,-2\ell;\dfrac{2r}{r+iL}\Big)\nonumber\\&&+C_2 F\Big(\ell+\frac{1}{2},\ell+1+\omega L,2\ell+2;\dfrac{2r}{r+iL}\Big)\Big]\;,\label{AdSsol1}
\end{eqnarray}
where $F(a, b, c; z)$ is the hypergeometric function, $C_1$ and $C_2$ are two integration constants.
\\
By imposing the first boundary condition in Eq.~\eqref{bc1},
one obtains the first relation between $C_1$ and $C_2$
\begin{equation}
\dfrac{C_1}{C_2}=2^{2\ell+1}\dfrac{\ell}{\ell-\omega L}\dfrac{\mathcal{A}_1}{F(\frac{1}{2}-\ell,1-\ell+\omega L,1-2\ell;2)}\;,\label{c1c2rel1}
\end{equation}
while by imposing the second boundary condition in Eq.~\eqref{bc2}, one obtains the second relation between $C_1$ and $C_2$
\begin{equation}
\dfrac{C_1}{C_2}=2^{2\ell+1}\ell\dfrac{\ell+1+\omega L}{\ell+1}\dfrac{F(\frac{3}{2}+\ell,2+\ell+\omega L,2\ell+3;2)}{\mathcal{A}_2}\;,\label{c1c2rel2}
\end{equation}
where
\begin{align}
\mathcal{A}_1=\;&F(\ell+\frac{1}{2},\ell+1+\omega L,2\ell+2;2)\nonumber\\&+F\left(\ell+\frac{3}{2},\ell+1+\omega L,2\ell+2;2\right)\;,\nonumber\\
\mathcal{A}_2=\;&2\ell F\left(-\ell-\frac{1}{2},-\ell+\omega L,-2\ell;2\right)\nonumber\\&+(\ell-\omega L) F\left(-\ell+\frac{1}{2},-\ell+1+\omega L,1-2\ell;2\right)\;.\nonumber
\end{align}
Then, expanding $R_1$ in Eq.~\eqref{AdSsol1} at small $r$, one gets
\begin{equation}
R_1\;\sim\;C_1\left(\frac{iL}{2}\right)^{2\ell+1}r^{-\ell-\frac{1}{2}}+C_2r^{\ell+\frac{1}{2}}\;.\label{AdSsol1origin}
\end{equation}
By requiring the solution to be regular at the origin, from Eq.~\eqref{AdSsol1origin}, one has to set $C_1=0$, which gives
\begin{eqnarray}
&&\mathcal{A}_1=0\;\nonumber\\&&\Rightarrow\;\;\omega_{1,N}L=2N+\ell+1\;,\label{normalmodes1}\\
&&F(\frac{3}{2}+\ell,2+\ell+\omega L,2\ell+3;2)=0\;\nonumber\\
&&\Rightarrow\;\;\omega_{2,N}L=2N+\ell+2\;,\label{normalmodes2}
\end{eqnarray}
where $N=0,1,2,\cdot\cdot\cdot$, and $\ell=\tfrac{1}{2},\tfrac{3}{2},\cdot\cdot\cdot$.

These are the Dirac normal modes on empty AdS. They have been previously derived in, $e.g.$~\cite{Cotaescu:1998ts}, wherein the Dirac equations were written in a specific form and a Cartesian gauge, by requiring the Dirac eigenfunctions to be regular at the boundary. Thus, the physical principle of vanishing energy flux boundary conditions is able to recover these results. It is worthwhile to note that the Dirac normal modes can \textit{not} be obtained by solving
Eq~\eqref{DiracU2radialR1} (or equivalently Eq.~\eqref{radialeq}) on an empty AdS spacetime with the commonly used scalar-like boundary condition.

We observe that the Dirac normal modes have the same expressions as in the Maxwell~\cite{PhysRevD.92.124006} and gravitational cases~\cite{Cardoso:2013pza} (see also~\cite{Natario:2004jd}). Similarly to these cases, the two sets of the Dirac normal modes are isospectral, up to one mode. In the numerical calculations, we are going to show that, the isospectrality will be broken when a BH is introduced in the bulk of the AdS spacetime.

\subsection{Analytic matching calculations for small Schwarzschild-AdS}
In this subsection, we perform an analytic calculation of quasinormal frequencies for a Dirac field on a Schwarzschild-AdS BH, with the two Robin boundary conditions given in Eqs.~\eqref{bc1} and~\eqref{bc2}. Such calculations are only valid for small Schwarzschild-AdS BHs ($r_+\ll L$), in the low frequency limit.

Following the well-known matching procedure, we shall  divide the region outside the event horizon into two sub-regions: the \textit{near region}, defined by the condition $r-r_+\ll1/\omega$, and the \textit{far region}, defined by the condition $r_+\ll r-r_+$. Then, we further require the condition $r_+\ll1/\omega$, so that an overlapping region exists wherein solutions obtained in the near region and in the far region are both valid. In the following analysis we focus on small AdS BHs, which allows us to solve the frequencies perturbatively, as deviations from the AdS normal modes.

\subsubsection{Near region solution}
In the near region, under the small BH approximation ($r_+\ll L$), it is convenient to define a new dimensionless variable
\begin{equation}
z\equiv 1-\dfrac{r_+}{r}\;,\nonumber
\end{equation}
to transform Eq.~\eqref{DiracU2radialR1} into
\begin{equation}
z(1-z)\dfrac{d^2R_1}{dz^2}+\dfrac{1-3z}{2}\dfrac{dR_1}{dz}+\left(\hat{\omega}\dfrac{1-z}{z}-\dfrac{k^2}{1-z}\right)R_1=0 \;,\label{neareq1}
\end{equation}
with
\begin{equation}
\hat{\omega}\equiv \left(\omega r_++\dfrac{i}{4}\right)^2+\dfrac{1}{16}\;.\nonumber
\end{equation}
The above equation can be solved in terms of the hypergeometric function
\begin{equation}
R_1\sim z^{\frac{1}{2}-i\omega r_+}(1-z)^{\ell+\frac{1}{2}}\;F(a,b,c;z)\;,\label{nearsol}
\end{equation}
with
\begin{equation}
a=\ell+1\;,\;\;\;
b=\ell+\frac{3}{2}-2i\omega r_+\;,\;\;\;
c=\frac{3}{2}-2i\omega r_+\;,\nonumber
\end{equation}
where an ingoing boundary condition at $r=r_+$ has been imposed.

In order to perform the matching with the far region solution below, the near region solution, Eq.~\eqref{nearsol}, should be expanded for large $r$. To achieve this, by taking the $z\rightarrow1$ limit and using the properties of the hypergeometric function~\cite{abramowitz+stegun}, we obtain
\begin{equation}
R_1 \;\sim \; \Gamma(c)\left[\dfrac{R^{\rm near}_{1,1/r}}{r^{\ell+\frac{1}{2}}} +R^{\rm near}_{1,r} r^{\ell+\frac{1}{2}}\right]
\;,\label{nearsolfar}
\end{equation}
where
\begin{align}
R^{\rm near}_{1,1/r} & \equiv \dfrac{\Gamma(-2\ell-1) r_+^{\ell+\frac{1}{2}}}{\Gamma(-\ell)\Gamma(\frac{1}{2}-\ell-2i\omega r_+)} \ ,\nonumber \\
R^{\rm near}_{1,r} &\equiv \dfrac{\Gamma(2\ell+1)r_+^{-\ell-\frac{1}{2}}}{\Gamma(\ell+1)\Gamma(\ell+\frac{3}{2}-2i\omega r_+)} \ .
\end{align}

%
\subsubsection{Far region solution}
In the far region, the BH effects can be neglected ($M\rightarrow0$), so that the solution for Eq.~\eqref{DiracU2radialR1} is the same as for the empty AdS spacetime, Eq.~\eqref{AdSsol1}. The two undetermined constants in Eq.~\eqref{AdSsol1} are related to each other by Eqs.~\eqref{c1c2rel1}--\eqref{c1c2rel2}, in order to satisfy the boundary conditions.

In order to match this solution with the near region solution, we expand Eq.~\eqref{AdSsol1} for small $r$, and obtain
\begin{equation}
R_1\;\sim\;\dfrac{R^{\rm far}_{1,1/r}}{r^{\ell+\frac{1}{2}}}+R^{\rm far}_{1,r}r^{\ell+\frac{1}{2}}\;,\label{farsolnear}
\end{equation}
with
\begin{eqnarray}
&&R^{\rm far}_{1,1/r}  \equiv 2^{-2\ell-1} (iL)^{2\ell+1} C_1 \ ,\nonumber \\
&&R^{\rm far}_{1,r} \equiv  C_2\;.\nonumber
\end{eqnarray}

\subsubsection{Overlap region}
To match the near region solution Eq.~\eqref{nearsolfar} and the far region solution Eq.~\eqref{farsolnear} in the intermediate region, we impose the matching condition $R^{\rm near}_{1,r}R^{\rm far}_{1,1/r}=R^{\rm far}_{1,r}R^{\rm near}_{1,1/r}$. Then we get
\begin{align}
&\dfrac{\Gamma(\ell+1)}{\Gamma(2\ell+1)}\dfrac{\Gamma(\ell+\frac{3}{2}-2i\omega r_+)}{\Gamma(-\ell+\frac{1}{2}-2i\omega r_+)}\dfrac{\Gamma(-2\ell-1)}{\Gamma(-\ell)}\left(\dfrac{r_+}{L}\right)^{2\ell+1}\nonumber\\
&=\left(\dfrac{i}{2}\right)^{2\ell+1}\dfrac{C_1}{C_2}\;.\label{matching}
\end{align}
Given the relations between $C_1$ and $C_2$, Eq.~\eqref{matching} becomes
\begin{align}
&\dfrac{\Gamma(\ell+1)}{\Gamma(2\ell+1)}\dfrac{\Gamma(\ell+\frac{3}{2}-2i\omega r_+)}{\Gamma(-\ell+\frac{1}{2}-2i\omega r_+)}\dfrac{\Gamma(-2\ell-1)}{\Gamma(-\ell)}\left(\dfrac{r_+}{L}\right)^{2\ell+1}\nonumber\\
&=i^{2\ell+1}\dfrac{\ell}{\ell-\omega L}\dfrac{\mathcal{A}_1}{F(\frac{1}{2}-\ell,1-\ell+\omega L,1-2\ell;2)}\;,\label{match1}
\end{align}
for the first boundary condition given by Eq.~\eqref{bc1}, and
\begin{align}
&\dfrac{\Gamma(\ell+1)}{\Gamma(2\ell+1)}\dfrac{\Gamma(\ell+\frac{3}{2}-2i\omega r_+)}{\Gamma(-\ell+\frac{1}{2}-2i\omega r_+)}\dfrac{\Gamma(-2\ell-1)}{\Gamma(-\ell)}\left(\dfrac{r_+}{L}\right)^{2\ell+1}\nonumber\\
&=i^{2\ell+1}\ell\dfrac{\ell+1+\omega L}{\ell+1}\dfrac{F(\frac{3}{2}+\ell,2+\ell+\omega L,2\ell+3;2)}{\mathcal{A}_2}\;,\label{match2}
\end{align}
for the second boundary condition given by Eq.~\eqref{bc2}.

Both Eqs.~\eqref{match1} and~\eqref{match2} can be solved perturbatively around the normal mode solutions, to obtain the imaginary part of quasinormal frequencies, in the small BH approximation. For a small BH, the left term in Eqs.~\eqref{match1} and~\eqref{match2} vanishes at the leading order, and we get the normal modes in an empty AdS spacetime, given by Eqs.~\eqref{normalmodes1} and~\eqref{normalmodes2}.

When the BH effects are taken into account, a correction to the frequency will be introduced
\begin{equation}
\omega_j L=\omega_{j,N} L+i\delta_j\;,\label{normalmode}
\end{equation}
where $j=1,2$ for the two different boundary \mbox{conditions}, and $\delta$ is used to describe the damping ($i.e.$ the imaginary part) of the quasinormal modes frequency. Replacing $\omega L$ in the second line of Eqs.~\eqref{match1} and~\eqref{match2} by $\omega_1 L$ and $\omega_2 L$ as given by \mbox{Eq.~\eqref{normalmode}}, we can obtain $\delta_j$ perturbatively, in terms of $r_+/L$.

Since the general expression for $\delta_j$ is quite messy, we only analyze Eqs.~\eqref{match1} and~\eqref{match2} for a subset of concrete values of the parameters. For the case with $\ell=\tfrac{1}{2}$ and $N=0$, from Eq.~\eqref{match1}, we get
\begin{equation}
\delta_1=-\dfrac{1}{4\pi}\dfrac{r_+^2}{L^2}+\mathcal{O}\left(\frac{r_+^2}{L^2}\right)\;,\label{analy1}
\end{equation}
and from Eq.~\eqref{match2}, we get
\begin{equation}
\delta_2=-\dfrac{3}{4\pi}\dfrac{r_+^2}{L^2}+\mathcal{O}\left(\frac{r_+^2}{L^2}\right)\;.\label{analy2}
\end{equation}
\\
Furthermore, by analyzing several cases with different $\ell$, we observe that,
\begin{equation}
-\delta_j\;\propto\;r_+^{2\ell+1}\;,\nonumber
\end{equation}
for both boundary conditions, while for bosonic fields the damping behaves as~\cite{PhysRevD.92.124006,Berti:2009wx}
\begin{equation}
-\delta_j\;\propto\;r_+^{2\ell+2}\;.\nonumber
\end{equation}

These analytic calculations may be used not only as the initial guess in a numerical procedure ($cf.$ next section) but also to double check the numerical results that we shall now address.

\section{Numerics}
\label{secnum}
In this part, we look for quasinormal frequencies for Dirac fields on Schwarzschild-AdS BHs \textit{numerically} by applying the vanishing energy flux boundary \mbox{conditions}, \textit{cf.} Eqs.~\eqref{bc1} and~\eqref{bc2}. We shall first briefly introduce the numerical methods employed, and then illustrate their application with some concrete examples.

\subsection{Method}
The numerical methods that we have employed to look for the characteristic eigenfrequency $\omega$ are of two types: a direct integration method, and the Horowitz-Hubeny method. The former works better for BHs with small size, while the latter works better for large BHs. As a consistency check, we find excellent agreement between these two methods when both are applicable.

\subsubsection{Direct integration approach}

To solve the radial equation~\eqref{DiracU2radialR1}, we may use the direct integration method, adapted from our previous works~\cite{Herdeiro:2011uu,Wang:2012tk,Wang:2014eha,Wang:2015fgp}. Firstly, we use Frobenius' method to expand $R_{1}$ close to the event horizon
\begin{equation}
R_{1}=(r-r_+)^\rho \sum_{j=0}^\infty c_j\;(r-r_+)^j\;,\nonumber
\end{equation}
with
\begin{equation}
\rho=\dfrac{1}{2}-\dfrac{i\omega r_+}{1+3r_+^2}\;,\nonumber
\end{equation}
to initialize Eq.~\eqref{DiracU2radialR1}, where the ingoing boundary condition at the horizon has been imposed, and the series expansion coefficients $c_j$ can be directly extracted after inserting these expansions into Eq.~\eqref{DiracU2radialR1}.

The asymptotic behavior of $R_{1}$ at infinity has been given in Eq.~\eqref{asysolR1}, where two coefficients, $\alpha_1$ and $\beta_1$, can be extracted from $R_1$ and its first derivative. For that purpose, we define two new fields $\left\{\chi,\psi\right\}$, which will asymptote respectively to $\left\{\alpha_1,\beta_1\right\}$, at infinity. Such a transformation can be written in matrix form by defining the vector $\mathbf{\Psi}^T=(\chi,\psi)$ for the new fields, and another vector $\mathbf{V}^T=(R_1,\frac{d}{dr}R_1)$ for the original field and its derivative. Then the transformation is given in terms of an $r$-dependent matrix $\mathbf{T}$ defined through
\begin{equation}
\mathbf{V}= \left(\begin{array}{cc} 1 & \frac{1}{r} \vspace{2mm}\\ 0 & -\frac{1}{r^2} \end{array}\right) \mathbf{\Psi} \equiv \mathbf{T} \mathbf{\Psi} \;.\nonumber
\end{equation}
To obtain a first order system of ODE for the new fields, we first define a matrix $\mathbf{X}$ through
\begin{equation}
\dfrac{d\mathbf{V}}{dr}=\mathbf{X}\mathbf{V} \; ,
\end{equation}
which can be read out from the original radial equation~\eqref{DiracU2radialR1}.
Then we obtain
\begin{equation}
\dfrac{d\mathbf{\Psi}}{dr}=\mathbf{T}^{-1}\left(\mathbf{X}\mathbf{T}-\dfrac{d\mathbf{T}}{dr}\right) \mathbf{\Psi} \;.\label{radialmatrix}
\end{equation}

\subsubsection{Horowitz-Hubeny approach}

The other method to solve for quasinormal frequencies for asymptotically AdS BHs, is the Horowitz-Hubeny approach~\cite{Horowitz:1999jd}. In order to employ this method, we first rewrite Eq.~\eqref{DiracU2radialR1} into the Schrodinger-like form
\begin{equation}
\dfrac{d^2\phi_1}{dr_\ast^2}+(\omega^2-V)\phi_1=0\;,\label{Schroeq}
\end{equation}
with
\begin{align}
V=&\dfrac{k^2\Delta_r}{r^4}-\dfrac{2\Delta_r^2}{r^6}+\dfrac{\Delta_r\Delta_r^\prime}{r^5}+\dfrac{\Delta_r^{\prime2}}{16r^4}-\dfrac{\Delta_r\Delta_r^{\prime\prime}}{4r^4}\nonumber\\
&-\dfrac{i\omega r^2}{2}\dfrac{d}{dr}\left(\dfrac{\Delta_r}{r^4}\right)\;,
\end{align}
where
\begin{equation}
\phi_1=\dfrac{r}{\Delta_r^{1/4}}R_1\; ;\label{transf0}
\end{equation}
the tortoise coordinate $r_\ast$ is defined as
\begin{equation}
\dfrac{dr_\ast}{dr}=\dfrac{r^2}{\Delta_r}\;,\nonumber
\end{equation}
and $\prime$ denotes derivative with respect to $r$.

By analyzing the near horizon behavior for $\phi_1$ in Eq.\eqref{Schroeq}, we find
\begin{equation}
\phi_1\thicksim e^{\pm i\bar{\omega}r_\ast}\;,\nonumber
\end{equation}
where
\begin{equation}
\bar{\omega}=\omega+\dfrac{i}{4r_+}\left(1+\dfrac{3r_+^2}{L^2}\right)\;.\label{omegabar}
\end{equation}
Then choosing the ingoing boundary condition and making the transformation
\begin{equation}
\phi_1=e^{-i\bar{\omega}r_\ast}\Phi_1\;,\label{transf}
\end{equation}
Eq.\eqref{Schroeq} may be rewritten as
\begin{equation}
S(x)\dfrac{d^2\Phi_1}{dx^2}+\dfrac{T(x)}{x-x_+}\dfrac{d\Phi_1}{dx}+\dfrac{U(x)}{(x-x_+)^2}\Phi_1=0\;,\label{HHeq}
\end{equation}
where we change variable $x=1/r$, in order to map the entire space outside the event horizon $r_+<r<\infty$ into a finite region $0<x<x_+$, with
\begin{align}
S(x)=&\left(c_0x^2+c_1x+c_2\right)^2\;,\nonumber\\
T(x)=&\left(c_0x^2+c_1x+c_2\right)\left(3c_0x^2-2x-2i\bar{\omega}\right)\;,\nonumber\\
U(x)=&\;\omega^2-\bar{\omega}^2+k^2\left(c_0x^3-x^2-1\right)+\dfrac{1}{2}i\omega\left(3c_0x^2-2x\right)\nonumber\\&-\dfrac{1}{16}\left(-15c_0^2x^4+20c_0x^3-4x^2+24c_0x-8\right)\;,
\label{Seriesfunc}
\end{align}
where
\begin{equation}
c_0=\dfrac{1+x_+^2}{x_+^3}\;,\;\;\;c_1=\dfrac{1}{x_+^2}\;,\;\;\;c_2=\dfrac{1}{x_+}\;,\nonumber
\end{equation}
and where $x_+=1/r_+$.

To evaluate quasinormal modes by using Horowitz-Hubeny approach, we expand all functions around $x_+$,
\begin{align}
&S(x)=\sum_{n=0}^4 s_n(x-x_+)^n\;,\nonumber\\
&T(x)=\sum_{n=0}^4 t_n(x-x_+)^n\;,\nonumber\\
&U(x)=\sum_{n=0}^4 u_n(x-x_+)^n\;,\label{funcsexp}
\end{align}
where the expansion coefficients $\{s_n, t_n, u_n\}$ can be read off from Eq.~\eqref{Seriesfunc}, and
\begin{equation}
\Phi_1=(x-x_+)^{\hat{\rho}}\sum_{j=0}^\infty a_j(x-x_+)^j\;.\label{phiexp}
\end{equation}
The index $\hat{\rho}$ and recurrence relations between $a_j$ can be obtained by substituting expansions in Eqs.~\eqref{funcsexp} and~\eqref{phiexp} into Eq.~\eqref{HHeq}. At the lowest order, the index $\hat{\rho}$ can be found as
\begin{equation}
\hat{\rho}=0\;,\;\;\;\;\;\;\hat{\rho}=-\dfrac{1}{2}+\dfrac{2i\omega x_+}{3+x_+^2}\;.\nonumber
\end{equation}
Since the ingoing boundary condition has been imposed in Eq.~\eqref{transf}, here we fix $\hat{\rho}=0$.
By comparing the other orders, we obtain recurrence relations
\begin{equation}
a_j=-\dfrac{1}{D_j}\sum_{n=1}^j[s_n(j-n)(j-n-1)+t_n(j-n)+u_n]a_{j-n}\;,\label{recurrrelation}
\end{equation}
with
\begin{equation}
D_j=s_0j(j-1)+t_0j+u_0\;.\nonumber
\end{equation}
From Eqs.~\eqref{transf0} and~\eqref{transf}, the boundary conditions in Eqs.~\eqref{bc1} and~\eqref{bc2} are now transformed into
\begin{equation}
\sum_ja_j(-x_+)^j\left(1+\dfrac{j}{\gamma x_+}\right)=0\;,\label{BCHH}
\end{equation}
with
\begin{equation}
\gamma=\gamma_1\equiv i\left(\ell+\frac{1}{2}-(\omega+\bar{\omega})\right)\;,\label{HHbc1}
\end{equation}
for the first boundary condition, and
\begin{equation}
\gamma=\gamma_2\equiv-i\left(\ell+\frac{1}{2}+(\omega+\bar{\omega})\right)\;,\label{HHbc2}
\end{equation}
for the second boundary condition, where $\bar{\omega}$ is given in Eq.~\eqref{omegabar}.

\subsection{Results}
When the BH size exceeds the parameter region where the analytic study is valid, quasinormal modes can only be solved numerically. In this part, we are going to present numerical results for Dirac quasinormal frequencies of Schwarzschild-AdS BHs, by employing the numerical methods described in the last subsection.

Before we exhibit our results, a couple of remarks are in order. In the numerical calculations all physical quantities are normalized by the AdS radius $L$ and we set $L=1$. Furthermore, we use $\omega_1$ ($\omega_2$) to represent the quasinormal frequency corresponding to the first (second) boundary condition.

In Table~\ref{DQNMs1}, we list a few fundamental $(N=0)$ quasinormal frequencies of $\omega_1$ (with $\ell=3/2$) and $\omega_2$ (with $\ell=1/2$), for different BH sizes. As we mentioned in the last section, the normal modes presented in Eqs.~\eqref{normalmodes1} and~\eqref{normalmodes2}, are isospectral under the mapping
\begin{equation}
\ell_1\leftrightarrow \ell_2+1 \ ,
\end{equation}
except one mode for $\omega_1$, where $\ell_1$ and $\ell_2$ refer to the angular momentum quantum number in the spectrum of $\omega_1$ and $\omega_2$. As one may observe from this table,  the presence of a BH  breaks the isospectrality. Note that such breakdown of the isospectrality occurs for all BH sizes, in particular for large BHs, which is in contrast to what occurs for the Maxwell case~\cite{PhysRevD.92.124006}.

To illustrate the difference between the two sets of modes, we present a few fundamental modes $(N=0)$ for $\omega_1$ and $\omega_2$ with the same $\ell$ $(\ell=1/2)$, in Table~\ref{DQNMs2}. As one may observe, for both modes, the real part of quasinormal frequencies first decreases then increases when increasing the BH size, while the magnitude of the imaginary part of quasinormal frequencies always increases. For a fixed BH size $r_+$, the magnitude of the imaginary part of $\omega_2$ is always larger than its counterpart of $\omega_1$. This implies that the first set of modes dominates the late time evolution of the interacting system.

For large BHs, as shown from Table~\ref{DQNMs1} and Table~\ref{DQNMs2}, the real part for either set of quasinormal modes varies slowly with the BH size (say from $r_+=50$ to $r_+=100$), while the imaginary part for both modes scales linearly with the BH size. This scaling law can be equally stated in terms of the Hawking temperature, which relates to the BH size through $T_H=3r_+/(4\pi L^2)$ for large BHs. We remark that this behavior is quite different from the one observed for the scalar case~\cite{Horowitz:1999jd}, for which both the real and imaginary parts scale linearly with the BH size. In Table~\ref{DQNMsNrp100}, we list a tower of quasinormal modes with different overtone numbers, $N$, for a BH with $r_+=100$. As one may observe, the excited modes $(N\geq1)$ for both sets are approximately evenly spaced in $N$.

\begin{figure*}
\begin{center}
\begin{tabular}{c}
\includegraphics[clip=false,width=0.396\textwidth]{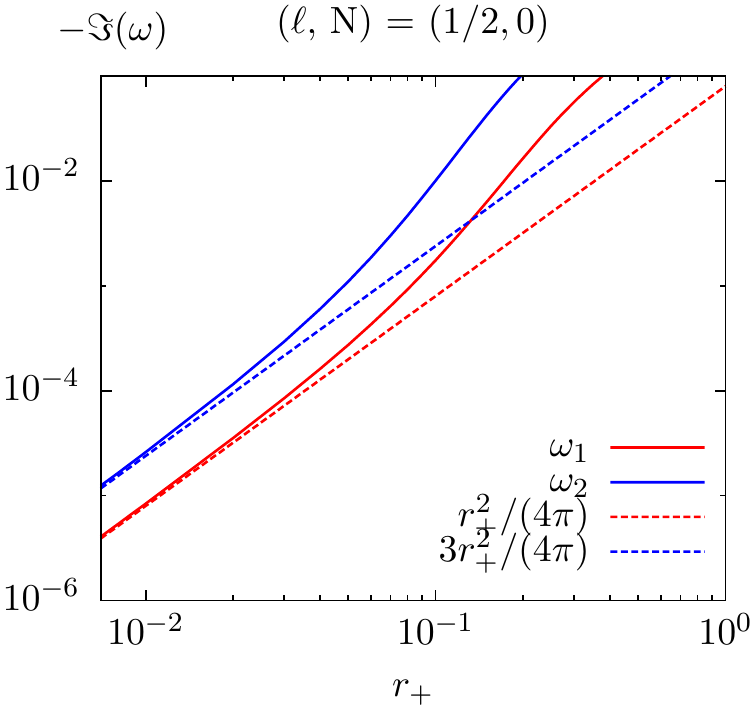}\;\;\;\;\;\;\;\;\;\;\;\;\;\;\;\;\;\;\;\;\;\;\;\includegraphics[clip=false,width=0.363\textwidth]{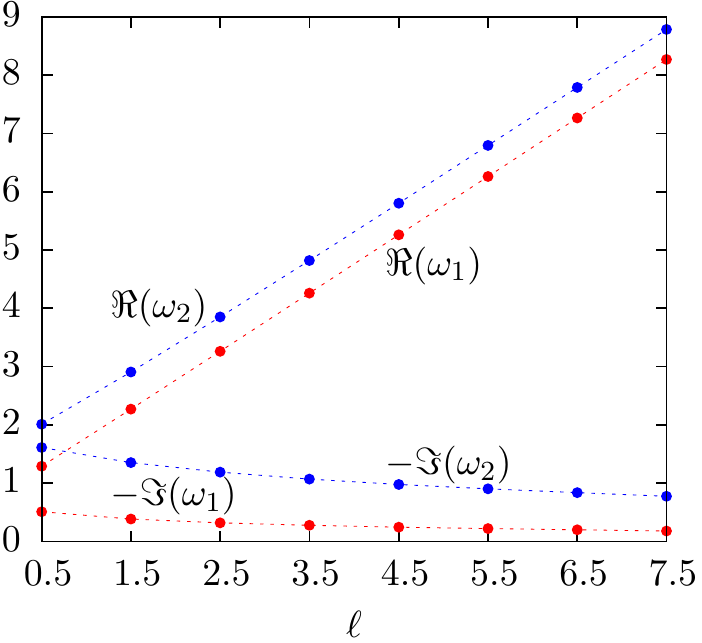}
\end{tabular}
\end{center}
\caption{\label{Df} (color online). Left: comparison of the imaginary part of quasinormal frequencies for the fundamental modes of each branch of solutions, between the analytic matching approximation for small BHs (dashed lines) and the numerical data (solid lines). Right: effects of the angular momentum quantum number $\ell$ on the quasinormal frequencies for intermediate BHs with $r_+=1$, and $N=0$.}
\end{figure*}

For small BHs, as shown in Section~\ref{secana} by an analytic matching method, the real part of the frequencies for both modes approach to the corresponding normal modes on empty AdS~\cite{Konoplya:2002zu}, given by Eqs.~\eqref{normalmodes1} and~\eqref{normalmodes2}, while the imaginary part of the frequencies approach to zero as
\begin{equation}
-\Im{(\omega)} \propto r_+^{2\ell+1}\;,\nonumber
\end{equation}
which behaviors differently with bosonic fields. Furthermore, we present a comparison between analytic results and numeric data in the left panel of Fig.~\ref{Df}, and find a good agreement for small $r_+$, which may be used to verify the validity of the analytic calculations but also another check for our numeric methods. In Table~\ref{DQNMsNrp01}, we list a tower of quasinormal modes with different overtone numbers $N$ for a BH with $r_+=0.1$. Observing the excited modes, we find they are also approximately evenly spaced in $N$.

\begin{table}
\caption{\label{DQNMs1} Two sets of quasinormal frequencies of fundamental modes for Dirac fields, for different BH size $r_+$ and $\ell$.}
\begin{ruledtabular}
\begin{tabular}{ l l l }
$r_+$ & $\omega_1(\ell=3/2)$ & $\omega_2(\ell=1/2)$ \\
\hline
0 & 2.5 & 2.5\\
0.2 & 2.4481 - 4.2096$\times 10^{-4}$ i & 2.1699 - 0.1041 i\\
0.5 & 2.3165 - 8.3949$\times 10^{-2}$ i & 1.9346 - 0.6738 i\\
0.8 & 2.2689 - 0.2647 i & 1.9530 - 1.2424 i\\
1.0 & 2.2708 - 0.3861 i & 2.0125 - 1.6134 i\\
5.0 & 2.8083 - 3.0906 i & 3.6912 - 9.5907 i\\
10 & 3.1965 - 6.9125 i & 4.9615 - 20.506 i\\
50 & 3.5619 - 37.326 i & 7.3070 - 111.38 i\\
100 & 3.5808 - 74.911 i & 7.5852 - 224.39 i\\
\end{tabular}
\end{ruledtabular}
\end{table}

\begin{table}
\caption{\label{DQNMs2} Two sets of quasinormal frequencies of fundamental modes for Dirac fields, for different BH size $r_+$ but the same $\ell$.}
\begin{ruledtabular}
\begin{tabular}{ l l l }
$r_+$ & $\omega_1(\ell=1/2)$ & $\omega_2(\ell=1/2)$ \\
\hline
0 & 1.5 & 2.5\\
0.2 & 1.4124 - 1.6293$\times 10^{-2}$ i & 2.1699 - 0.1041 i\\
0.5 & 1.3007 - 0.1784 i & 1.9346 - 0.6738 i\\
0.8 & 1.2836 - 0.3789 i & 1.9530 - 1.2424 i\\
1.0 & 1.2930 - 0.5130 i & 2.0125 - 1.6134 i\\
5.0 & 1.6155 - 3.4826 i & 3.6912 - 9.5907 i\\
10 & 1.7302 - 7.3217 i & 4.9615 - 20.506 i\\
50 & 1.7907 - 37.459 i & 7.3070 - 111.38 i\\
100 & 1.7929 - 74.980 i & 7.5852 - 224.39 i\\
\end{tabular}
\end{ruledtabular}
\end{table}

For intermediate BHs, it seems that the real part for both quasinormal modes reaches a minimum. In Table~\ref{DQNMsNrp1}, we list a tower of quasinormal modes with different overtone numbers $N$ for a BH with $r_+=1$. We observe that the excited modes are again approximately evenly spaced in $N$. Moreover, we consider BHs with $r_+=1$ to exemplify the effect of the angular momentum quantum number $\ell$ on both frequencies. As one may see from the right panel of Fig.~\ref{Df}, for both modes, the real (imaginary) part of quasinormal frequencies increases (decreases) in magnitude as $\ell$ increases. This behavior is qualitatively similar for other BH sizes.

\begin{table}
\caption{\label{DQNMsNrp100} Quasinormal frequencies of the Dirac field on Schwarzschild-AdS BHs with $r_+=100$, $\ell=1/2$ and different overtone number $N$.}
\begin{ruledtabular}
\begin{tabular}{ l l l }
$N$ & $\omega_1$ & $\omega_2$ \\
\hline
0 & 1.7929 - 74.980 i & 7.5852 - 224.39 i\\
1 & 32.379 - 364.31 i & 83.299 - 480.41 i\\
2 & 146.19 - 593.52 i & 209.76 - 707.40 i\\
3 & 273.44 - 821.03 i & 337.27 - 934.48 i\\
4 & 401.23 - 1047.8 i & 465.28 - 1161.0 i\\
5 & 529.42 - 1274.1 i & 593.62 - 1387.2 i\\
6 & 657.89 - 1500.2 i & 722.20 - 1613.1 i\\
\end{tabular}
\end{ruledtabular}
\end{table}


\begin{table}
\caption{\label{DQNMsNrp01} Quasinormal frequencies of the Dirac field on Schwarzschild-AdS BHs with $r_+=0.1$, $\ell=1/2$ and different overtone number $N$.}
\begin{ruledtabular}
\begin{tabular}{ l l l }
$N$ & $\omega_1$ & $\omega_2$ \\
\hline
0 & 1.4629 - 1.7447$\times 10^{-3}$ i & 2.3557 - 1.0007$\times 10^{-2}$ i\\
1 & 3.1922 - 3.7362$\times 10^{-2}$ i & 3.9916 - 0.1021 i\\
2 & 4.7796 - 0.2131 i & 5.5800 - 0.3603 i\\
3 & 6.4004 - 0.5256 i & 7.2375 - 0.6970 i\\
4 & 8.0861 - 0.8693 i & 8.9424 - 1.0410 i\\
5 & 9.8041 - 1.2117 i & 10.670 - 1.3814 i\\
6 & 11.538 - 1.5502 i & 12.409 - 1.7182 i\\
\end{tabular}
\end{ruledtabular}
\end{table}


\begin{table}
\caption{\label{DQNMsNrp1} Quasinormal frequencies of the Dirac field on Schwarzschild-AdS BHs with $r_+=1$, $\ell=1/2$ and different overtone number $N$.}
\begin{ruledtabular}
\begin{tabular}{ l l l }
$N$ & $\omega_1$ & $\omega_2$ \\
\hline
0 & 1.2930 - 0.5130 i & 2.0126 - 1.6134 i\\
1 & 2.8897 - 2.8248 i & 3.8066 - 4.0244 i\\
2 & 4.7455 - 5.2213 i & 5.6954 - 6.4141 i\\
3 & 6.6525 - 7.6044 i & 7.6144 - 8.7925 i\\
4 & 8.5797 - 9.9791 i & 9.5476 - 11.164 i\\
5 & 10.518 - 12.349 i & 11.489 - 13.532 i\\
6 & 12.462 - 14.715 i & 13.436 - 15.897 i\\
\end{tabular}
\end{ruledtabular}
\end{table}

\section{Discussion and Final Remarks}
\label{discussion}
In this paper we have studied Dirac quasinormal modes on Schwarzschild-AdS BHs, from a \textit{new} perspective on the boundary condition. For this purpose we first derived the Dirac equations and constructed the energy flux for Dirac fields. Following the principle we proposed in~\cite{PhysRevD.92.124006} that \textit{the energy flux should vanish at the AdS boundary}, we obtained \textit{two} distinct sets of boundary conditions. These boundary conditions were then employed to calculate Dirac normal modes on empty AdS and quasinormal modes on Schwarzschild-AdS BHs.

On an empty AdS spacetime, we solved the Dirac equations \textit{analytically} and obtained two branches of normal modes, albeit isospectral, corresponding to the two sets of boundary conditions. This is an interesting result because Dirac normal modes on empty AdS can \textit{not} be obtained by solving Eq.~\eqref{DiracU2radialR1} (or equivalently Eq.~\eqref{radialeq}) when imposing the commonly used scalar-like boundary condition. We remark these two spectra have the exact same expressions as for the Maxwell~\cite{PhysRevD.92.124006} and gravitational~\cite{Dias:2013sdc} cases.

In the case of Schwarzschild-AdS, we used both analytic and numerical methods to study Dirac quasinormal modes. In the small BH limit, we obtained the imaginary part of the quasinormal frequencies by an analytic matching method, which shows explicitly how two branches of quasinormal modes emerge from two sets of boundary conditions. We found $-\Im(\omega)\propto r_+^{2\ell+1}$ for the two sets of boundary conditions. This behavior is different from that of the bosonic fields, for which $-\Im(\omega)\propto r_+^{2\ell+2}$. We then varied the BH size $r_+$, the angular momentum quantum number $\ell$, and the overtone number $N$ in the numeric calculations, and analyzed their effects on the two branches of Dirac quasinormal modes.

In a nutshell, one observed the following trends. The real part for both quasinormal modes first decreases and  then increases when increasing the BH size $r_+$, while the magnitude of the imaginary part for both quasinormal modes always increases. By increasing the angular momentum quantum number $\ell$, the real part for both modes increases roughly linearly, while the imaginary part decreases but varies weakly. Varying the overtone number $N$, we found that excited modes of both sets for all BH sizes are approximately evenly spaced in $N$. Furthermore, the first branch of modes dominate at the late time evolutions. Dirac quasinormal modes were also calculated for $R_2$ equation, and we obtained the exactly same results.

This framework can be applied not only to the $\gamma$ matrices formalism, but also to the Teukolsky formalism. For Dirac fields, these two formalisms are simply related by the transformations presented in Appendix~\ref{app1}. Then one may easily verify that two sets of boundary conditions can be obtained for the Teukolsky variables. These conditions lead to two branches of quasinormal modes, which are exactly the same as we reported in the above, from the $\gamma$ matrices formalism.

Our work shows the robustness of the ``vanishing energy flux" principle, to set boundary conditions, in the sense that they are applicable not only for bosonic fields but also for fermionic fields. To fully explore these new boundary conditions for a Dirac field, we are going to generalize the present work to calculate quasinormal modes of charged Dirac fields~\cite{chargedDirac} and on rotating background~\cite{rotatingDirac}, which will be hopefully reported soon.

A final remark goes for the Dirac number current. As we have shown in Appendix~\ref{app2}, requiring a vanishing number current leads to the same boundary conditions as requiring a vanishing energy flux. Physically this is clear since energy flux is equivalent to the number current up to the particle's energy. Technically, however, the number current is much easier to calculate. Therefore, the Dirac number current may be used as an alternative to the energy flux, to study Dirac quasinormal modes.

\bigskip

\noindent{\bf{\em Acknowledgements.}}
This work is supported by the National Natural Science Foundation of China under Grant No. 11705054. C.H. acknowledges funding from the FCT IF programme, and his work is also supported by the EU grants H2020-MSCA-RISE-2015 Grant No. StronGrHEP-690904, and by the CIDMA strategic project UID/MAT/04106/2013. J.Jing's work is partially supported by the National Natural Science Foundation of China under Grant No. 11475061.

\bigskip

\appendix

\section{Dirac equations in the Newman-Penrose formalism}
\label{app1}
In the Newman-Penrose formalism, following the celebrated work by Teukolsky~\cite{Teukolsky:1973ha}, the Dirac equations have already been derived in~\cite{Khanal:1983vb,Dias:2012pp}. In this appendix we rewrite these equation by adapting to our notation, describing a spin $s$ $(s=\pm1/2)$ perturbation.

The radial equation is
\begin{equation}
\Delta_r^{-s}\dfrac{d}{dr}\left(\Delta_r^{s+1}\dfrac{d R_{s}(r)}{dr}\right)+H(r)R_{s}(r)=0\;,\label{radialeq}
\end{equation}
with
\begin{equation}
H(r)=\dfrac{K_r^2-i s K_r \Delta_r^\prime}{\Delta_r}+2isK_r^\prime+\dfrac{s+|s|}{2}\Delta_r^{\prime\prime}
-k^2\;,\nonumber
\end{equation}
where
\begin{equation}
K_r=\omega r^2\;,\;\;\;\;\;\;k^2=(\ell+\tfrac{1}{2})^2\;,\nonumber
\end{equation}
while the angular equation is
\begin{equation}
\dfrac{1}{\sin\theta}\dfrac{d}{d\theta}\left(\sin\theta\dfrac{dS_{lm,s}}{d\theta}\right)+A(\theta)S_{lm,s}=0\;,\label{angulareq}
\end{equation}
with $S_{lm,s}\equiv S_{lm,s}(\theta)$, and where
\begin{equation}
A(\theta)=-\dfrac{m^2}{\sin^2\theta}-\dfrac{2ms\cos\theta}{\sin^2\theta}-s^2\cot^2\theta+\ell(\ell+1)-s^2\;.\nonumber
\end{equation}

Our purpose in presenting the Dirac equations in the Newman-Penrose formalism is to argue the \textit{universality} of the results we have obtained with the vanishing energy flux boundary condition. That is, these boundary conditions can be applied either to Eqs.~\eqref{DiracU2radialR1},~\eqref{DiracU2radialR2} or to Eq.~\eqref{radialeq}, yielding the same results. Comparing Eq.~\eqref{radialeq} with Eqs.~\eqref{DiracU2radialR1},~\eqref{DiracU2radialR2} and Eq.~\eqref{angulareq} with Eqs.~\eqref{DiracU2angularS1},~\eqref{DiracU2angularS2}, we observe that the two sets of radial equations are simply related by the transformations $R_1=R_{-1/2}, R_2=\sqrt{\Delta_r}R_{+1/2}$; while the two sets of angular equations are related by the transformations $S_1=\sqrt{\sin\theta}S_{-1/2}, S_2=\sqrt{\sin\theta}S_{+1/2}$. Based on this observation, it easily follows that the vanishing energy flux boundary conditions can be applied to the Teukolsky formalism.

\section{The number current of Dirac fields}
\label{app2}
In this appendix, we present the derivation of the number current for Dirac fields, and show explicitly that vanishing energy flux leads to vanishing number current.

The number current for Dirac fields is defined as
\begin{equation}
J^\mu=\bar{\Psi}\gamma^\mu\Psi\;,\label{current1}
\end{equation}
with $\bar{\Psi}=\Psi^\dag\gamma^0$, where $\Psi^\dag$ is the adjoint of $\Psi$.

The radial component of the number current is
\begin{equation}
J^r=\bar{\Psi}\gamma^r\Psi\;.\label{current2}
\end{equation}
Substituting the $\gamma$ matrices in Eq.~\eqref{gammams}, together with the field's decompositions in Eqs.~\eqref{psidecom1},~\eqref{psidecom2}, Eq.~\eqref{current2} becomes
\begin{equation}
J^r=\dfrac{2}{r^2\sin\theta}\left(|R_1|^2|S_1|^2-|R_2|^2|S_2|^2\right)\;.\nonumber
\end{equation}
Then integrating the current over a sphere, we obtain
\begin{equation}
\mathcal{J}|_r = \int_{S^2}\sin\theta d\theta d\varphi r^2J^r\propto ( |R_1|^2-|R_2|^2 ) \;,\label{currcon}
\end{equation}
where the normalization condition for $S_1$ and $S_2$ has been employed.

Comparing Eq.~\eqref{currcon} with Eq.~\eqref{flux2}, one concludes that vanishing energy flux leads to vanishing number current. Since the calculation for the number current is much easier, one may use the number current condition as an alternative to calculate Dirac quasinormal frequencies, in particular, on a more complicated background.

\bibliographystyle{h-physrev4}
\bibliography{DiracSAdS} 

\end{document}